\documentclass[letterpaper, 10 pt, conference]{ieeeconf}  

\IEEEoverridecommandlockouts                              

\usepackage{amsmath,amsthm}
\usepackage{amsmath,amssymb,amsfonts}
\usepackage{algorithmic}
\usepackage{graphicx}
\usepackage{textcomp}
\usepackage{listings}
\usepackage{multirow}
\usepackage{xcolor}
\usepackage{mathtools}
\usepackage{graphicx}
\usepackage{array}
\usepackage{cancel}
\usepackage[linesnumbered,algoruled,boxed,lined]{algorithm2e}
\usepackage{textcomp}
\usepackage{comment}

\usepackage{times}  
\usepackage{helvet}  
\usepackage{todonotes}
\usepackage{bm}
\usepackage{tikz}
\usetikzlibrary{arrows,automata}
\usepackage{float}
\usepackage{subcaption}
\usepackage{mathtools}
\usepackage{acronym}
\usepackage{courier}  
\usepackage{url}
\usepackage{colonequals}
\usepackage{graphicx}  
\usepackage{paralist}

\usetikzlibrary{shapes.geometric,fit}

\newif\ifuseboldmathops
\newif\ifuseittextabbrevs
\useboldmathopstrue   

\ifuseittextabbrevs

    \newcommand{\ie}{{\it i.e.}}

\else

    \newcommand{\ie}{i.e.}

\fi

\ifuseboldmathops

\else

\fi

\ifuseboldmathops

\else

\fi

\ifuseboldmathops
    \newcommand{\Expect}{\mathop{\bf E{}}\nolimits}
    
\else
    \newcommand{\Expect}{\mathop{\mathbb{E}{}}\nolimits}
    
\fi

\ifuseboldmathops

\else

\fi

\newcommand{\argmax}{\mathop{\mathrm{argmax}}}

\newcommand{\sink}{\mathsf{sink}}

\newcommand{\abs}[1]{\lvert#1\rvert}

\newcommand{\dist}[1]{\mathsf{Dist}(#1)}

\DeclareMathOperator{\topologies}{\Gamma}
\DeclareMathOperator{\init}{\gamma}

\newcommand{\calP}{\mathcal{P}}

\newcommand{\calF}{\mathcal{F}}

\newcommand{\qvalue}{\mathcal{Q}}

\DeclareMathOperator{\mdp}{M}
\DeclareMathOperator{\mdptrans}{\mathcal{P}}

\newcommand{\nat}{\mathbb{Z}}

\theoremstyle{definition}
\newtheorem{definition}{Definition}

\newtheorem{problem}{Problem}

\newtheorem{assumption}{Assumption}

\acrodef{mdp}[MDP]{Markov decision process}
\acrodef{ssp}[SSP]{Stochastic Shortest Path}
\acrodef{mtd}[MTD]{Moving Target Defense}
\acrodef{pddl}[PDDL]{Planning Domain Definition Language}
\acrodef{milp}[MILP]{Mixed-Integer Linear Programming}
\acrodef{lp}[LP]{Linear Programming}
\acrodef{cvss}[CVSS]{Common Vulnerability Scoring System}
\acrodef{os}[OS]{Operating System}

\DeclareMathOperator*{\optmins}{\textrm{min.}}

\DeclareMathOperator*{\optsts}{\textrm{s.t.}}

\title{\LARGE \bf
 Synthesis of Proactive Sensor Placement In Probabilistic Attack Graphs
}

\author{Lening Li$^{1}$, Haoxiang Ma$^2$, Shuo Han$^3$ and Jie Fu$^2$  
\thanks{*Research was sponsored by the Army Research Office and was accomplished under Grant
Number W911NF-22-1-0166.}
\thanks{$^{1}$Lening Li is with Department of Robotics Engineering, Worcester Polytechnic Institute, Worcester, MA 01609, USA {\tt\small lli4@wpi.edu}}%
\thanks{$^{2}$Haoxiang Ma, Jie Fu are with the Department of Electrical \& Computer Engineering, University of Florida, Gainesville, FL, 32611, USA {\tt\small hma2, fujie@ufl.edu}}%
\thanks{$^{3}$Shuo Han is with the Department of Electrical \& Computer Engineering, University of Illinois at Chicago, Chicago, IL 60607, USA
        {\tt\small hanshuo@uic.edu}}%
}

\begin{document}

\maketitle
\begin{abstract}
    This paper studies the deployment of joint moving target defense (MTD) and deception against multi-stage cyberattacks. Given the system equipped with MTD that randomizes between different  configurations, we investigate how to allocate a bounded number of sensors in each configuration to optimize the attack detection rate before the attacker achieves its objective. Specifically, two types of sensors are considered: intrusion detectors that are observable by the attacker and stealthy sensors that are not observable to the attacker. We propose a two-step optimization-based approach for allocating intrusion detectors and stealthy sensors: Firstly, the defender allocates intrusion detectors assuming the attacker will best respond to evade detection by intrusion detectors. Secondly, the defender will allocate stealthy sensors, given the best response attack strategy computed in the first step, to further reduce the attacker's chance of success. We illustrate the effectiveness of the proposed methods using a cyber defense example.
\end{abstract}

\thispagestyle{empty}
\pagestyle{empty}

\section{Introduction}
\label{sec:introduction}

This paper considers a game-theoretic design of a proactive cyber defense system using a combination of \ac{mtd}, intrusion detectors, and deception (with stealthy sensors). Proactive defense means that the defender does not know the attacker's presence or the progress made by the attack but employs randomization to thwart and mitigate attacks. For example, an attack action can fail if the system configuration changes and invalidates the targeted vulnerability. Meanwhile, the defender can deploy sensors to detect the attacker at the early stage of the attack. Nonetheless, with the increasingly advanced \ac{mtd}~\cite{zhuang2014towards}, detection, and cyber deception~\cite{jajodiaCyberDeceptionBuilding2016a}, it remains a challenge to assess the effectiveness of a combination of \ac{mtd} and sensor-based detection mechanisms, let alone to design an effective cyber defense system with joint \ac{mtd} and deception. In this paper, we integrate a formal-method modeling and optimization-based approaches to address the following question: ``how to allocate a limited number of (potentially heterogeneous) sensors in this system to maximize the probability of attack detection before that the attacker achieves its objective?'' ``What is the benefit of employing deceptive, stealthy sensors for proactive defense?''

To model the effects of \ac{mtd} on the attack performance, we employ a variant of attack graphs~\cite{jha2002two,hewett2008host}, which models the causal and logical dependencies between system's vulnerabilities or attacker's subgoals observed in multi-stage attacks. We introduce the Markov Chain as a formal model of a class of \ac{mtd} in which the defender switches randomly between different system configurations~\cite{islamspecificationdrivenmovingtarget2019a,al2012random,hongAssessingEffectivenessMoving2016}. Given a system equipped with such an \ac{mtd} strategy, we first capture the attacker's decision-making problem using an \ac{mdp} with a reachability objective; that is, the attacker aims to reach some goal states eventually while evading detection by sensors. For example, the attacker's goal state can be that the attacker gains root access to a critical database server. Then, we focus on the synthesis problem for the defender to minimize the attack success rate by optimally allocating two types of sensors: intrusion detectors that are observable by the attacker and stealthy sensors that are unobservable to the attacker. A stealthy sensor can be realized by honey patching~\cite{araujo2014patches} of a known vulnerability. When the attacker exploits a honey-patched vulnerability, he will be detected. We incorporate the attacker's safety constraints, \ie, evading sensor detection, into the attack objective and formulate a bi-level optimization problem. We then design \acp{milp}s in a two-step manner to approximately optimize intrusion detectors and stealthy sensors allocation for the defender.

\paragraph*{Related work}
The synthesis of proactive defense strategies studied herein is closely related to the Stackelberg security game (SSG) (surveyed in~\cite{sinhaStackelbergSecurityGames2018}). In an SSG, the defender is to defend a set of targets with limited resources, while the attacker selects the optimal attack strategy given the knowledge of the defender strategy. The solution concepts of Stackelberg Equilibrium are employed by~\cite{senguptaMovingTargetDefense2018a} to design a mixed strategy for the defender to allocate intrusion detectors and implement the intrusion detectors randomization schedule using \ac{mtd}. In~\cite{nguyenMultistageAttackGraph2018}, the authors formulate the security countermeasure-allocation problem as a resource-allocation game, where attack graphs are used to evaluate the security of the network given the allocated resources. A Bayesian attack graph is an empirical attack behavior model constructed from the data and exploitability of the targeted vulnerability~\cite{yoonAttackGraphbasedMoving2020}. In~\cite{miehlingOptimalDefensePolicies2015b}, the authors assume that a Bayesian attack graph~\cite{frigaultMeasuringNetworkSecurity2008a} represents the attacker's behavior and design optimal defender strategies under partial observations using solutions of partially observable Markov decision processes. Another related formulation is the plan interdiction problem studied in~\cite{letchfordOptimalInterdictionAttack}, where the attacker is to reach a subset of goals with attack actions, and the defender is to mitigate the attack by interdicting or removing the attack actions. They formulated a mixed-integer programming problem to maximize the defender's objective function assuming the optimal plan of the attacker given the interdiction strategy.

In comparison to existing work, we introduce a formal model of \ac{mtd} strategy and capture the effects of \ac{mtd} on system configuration randomization as a probabilistic switching between different attack graphs. For allocating intrusion detectors given a randomization schedule, we consider the optimal allocation given a ``worst-case'' attacker who knows about the \ac{mtd} schedule and the locations of intrusion detectors and plans to evade detection by intrusion detectors. In addition, we allocate stealthy sensors, which are unobservable to the attacker, to decrease the attack success rate further. To the best of our knowledge, the combined effect of \ac{mtd} and cyber deception has not been investigated in the literature. This work contributes a formal method-based approach for modeling and synthesizing approximately optimal cyber defense with a class of sensor deception.

\section{Problem Formulation}
\label{sec:problem_formulation}
Our modeling of the attack-defend interaction is inspired by the formal graphical security model called \emph{attack graphs}, introduced in~\cite{jhaTwoFormalAnalyses2002} for modeling sequential attacks in a network. Specifically, in network security, an attack graph is constructed from the attack actions (vulnerabilities in a program/network) and the pre- and post-conditions of actions.

Besides cybersecurity applications, attack graphs are commonly used for analyzing terrorist networks, counter-terrorism networks, and transportation networks (see a survey in~\cite{ionita2017graphical}). In this work, though the examples are set with cyber security applications in mind, similar solution approaches are applicable for general security problems modeled using attack graphs.
\begin{definition}[Attack Graph]
    Given a system configuration, the corresponding attack graph is represented as a probabilistic transition system $TS = \langle S, A, T, \nu_0, F \rangle$,
    where
    \begin{inparaenum}[1)]
        \item $S$ is a finite set of states, representing security-related attributes of the system and the attacker;
        \item $A$ is a finite set of attack actions;
        \item $T \colon S \times A \to \dist{S}$ is a probabilistic transition function that maps a state-action pair into a distribution over next states;
        \item $\nu_0$ is the initial state distribution;
        \item $F \subseteq S$ is a subset of states. The attacker's objective is to reach one of the states in $F$.
    \end{inparaenum}
\end{definition}
A \emph{path} $\rho = s_0 a_0 s_1 a_1 \ldots$ of $TS$ is a state-action sequence such that for any $i \ge 0$, there exist $a \in A$, for which $T(s_{i+1} \mid s_i, a)>0$.

In   cybersecurity, an example of a state can be ``the attack is at host 1, and host 2 running an \texttt{ftp} server''. An attack action can be to exploit a known vulnerability on the \texttt{ftp} server, to reach a state where  ``the attacker has user access to host 2.'' In relation to logical attack graph~\cite{ou2006scalable}, one can employ PDDL language~\cite{gerevini2009deterministic} to generate a (deterministic) transition system from the pre- and post-conditions of exploitation actions in logical attack graphs.


\noindent \textbf{Defender's proactive, randomized moves}
A defense configuration $i$ describes the network connectivity, node configurations, defensive countermeasures, and the allocation of sensors. It is observed that the changes in system configuration can be directly captured by the changes in the attack graph, including removing/adding transitions or changing the probability distributions given state-action pairs.

Let $\topologies$ be the set of indices of different system configurations among which the defender switches. Each configuration $i \in \topologies$ generates a probabilistic transition system  $TS^{i} = \langle S, A , T^i, \nu_0 , F \rangle$. To simplify notations, we assume that different configurations $i$ and $j$ will have different transition functions but share all other components. Note that if the transition systems constructed from two attack graphs of different configurations do not have the same set of states, then we can make the union of the state sets as the set $S$. The same argument applies to justify the same set of attack actions with different configurations.

Next, we introduce a computational model of proactive defense strategies using randomization.
\begin{definition}[Proactive Defense Strategy]
    A proactive defense strategy is defined by a Markov Chain
    \[
        MC = \langle \topologies, P, \init \rangle,
    \]
    where
    \begin{itemize}
        \item $\topologies$ is a finite set of system configurations.
        \item $P \colon \topologies \to \dist{\topologies}$ is the probabilistic transition function. Given the current configuration $i$, the probability of reaching configuration $j$ is $P(i,j)$.
        \item $\init$ is an initial state distribution of configurations.
    \end{itemize}
    \label{def:sch}
\end{definition}

\noindent \textbf{Defender's Proactive Intrusion Detection with Deceptive Sensors}
Besides randomization, the defender can allocate sensors to monitor different subsets of states. The defender can block the attacker from the network when a sensor detects an attack.

Specifically, we consider two kinds of sensors: the first kind, called \emph{intrusion detectors}, can be detectable by the attacker; and the second kind, called \emph{stealthy sensors}, cannot be detected by the attacker unless the attacker directly interacts with it. In practice, intrusion detectors are intrusion detection systems or firewalls. Stealthy sensors can be realized by honeypots and honey patching~\cite{araujo2014patches}. A honey patch misleads the attacker into believing a specific vulnerability exists. However, such a vulnerability is patched, and exploitation of it will be directly detected by the defender. Honey patching has been recently proposed as an effective detection mechanism using cyber deception.

\begin{definition}[Sensor Allocation]
    The defender's sensor allocation design is a pair of Boolean-valued vectors $(\vec{x}, \vec{y}) $, where $\vec{x}, \vec{y} \in \{0,1\}^{\abs{S\times \topologies \times A}}$ such that
    \begin{itemize}
        \item $\vec{x}_{s,i,a}=1$ if and only if under the configuration $i$, the intrusion detector is placed on state-action pair $(s,a)$.
        \item $\vec{y}_{s,i,a}=1$ if and only if under the configuration $i$, the stealthy sensor is placed on state-action pair $(s,a)$.
    \end{itemize}
    \label{def:sensor_allocation}
\end{definition}

It is observed that this modeling of the defender's observation captures realistic sensing modalities. For example, an intrusion detector may only be able to detect one type of exploitation/action from a given state. A similar argument applies to honey patching, which is used to detect the exploitation of a specific known vulnerability on a target system. Note that the action can still be detected even if the attack fails (with the probabilistic action outcomes).

\begin{assumption}[Sensor Allocation Constraints]
    \label{assume:sensor-constraints}
    For any configuration $i\in \topologies$, the intrusion detector can be allocated to a subset $I \subset S\times A$ of state-action pairs and the stealthy sensor can be allocated to a subset $H\subset S\times A$. The set $I$ and $H$ may have a nonempty intersection. However, if, for configuration $i \in \topologies$, $(s,a)$ is allocated with the intrusion detector, then it cannot be allocated with a stealthy sensor at the same configuration, and vice versa.
\end{assumption}

\begin{problem}
Consider the set of attack graphs $\{TS^i\mid i \in \topologies\}$ for different system configurations, the set of goal states $F$, and the defender's randomization schedule modeled as a Markov Chain $MC$. Assuming the sensor allocation constraints in Assumption~\ref{assume:sensor-constraints}, compute a sensor allocation strategy given a finite number $h\in \nat $ of intrusion detectors and $k\in \nat$ of stealthy sensors such that the defender can maximize the probability of detecting the attacker before the attacker reaches a goal state in $F$.
\end{problem}

\section{A Stackelberg Game Formulation}
To formulate the sensor allocation problem, we first construct a model that describes the attacker's interaction with the defense system using randomization but no sensors, then we show how a fixed sensor allocation $(\vec{x},\vec{y})$ can change such a model to different models perceived by the attacker and the defender.

\begin{assumption}
    It is assumed that the defender and attacker move concurrently. At every time step, the attacker selects an attack action and the defender makes a probabilistic move.
\end{assumption}

\begin{definition}[Attacker's Markov Decision Process without Sensors]
    \label{def:attackMDP}
    Given a proactive defense strategy $MC = \langle  \topologies, P, \init \rangle$, a set of probabilistic transitions systems $\{TS^i = \langle S, A, T^i, \nu_0 , F \rangle \mid i \in \topologies\}$ generated from different network configurations, the attacker's planning problem is captured by the \ac{mdp}:
    \[
        \mdp = \langle Z, A, \mdptrans, \iota, \calF \rangle
    \]
    with following components:
    \begin{itemize}
        \item $Z \colon S \times \topologies$ is the set of states.
        \item $A$ is the set of attack actions.
        \item $\mdptrans \colon Z \times A \to \dist{Z}$ is a probabilistic transition function defined as follows. Consider $(s, i), (s', j) \in Z$, for each action $a \in A$, we consider two cases:
              \begin{enumerate}[(a)]
                  \item If $T^j(s' \mid s,a)>0$, then $\mdptrans((s', j) \mid (s,i), a) = P(i,j)\cdot T^j(s' \mid s,a)$;
                  \item If $T^j(s' \mid s,a)$ is not defined, then we have $\mdptrans((s, j) \mid (s, i), a) = P(i,j) $. In this case, the defense state changes, but no progress is made by the attacker. This is because that the attack action is invalid given the updated configuration.
              \end{enumerate}
        \item $\iota$ is the initial state distribution, defined by the joint distribution of initial state distribution $\nu_0$ in the attack graph and the initial state distribution $\init$ of the proactive defense strategy.
        \item $\calF = F \times \topologies$ is the set of final states which the attacker is to reach.
    \end{itemize}
\end{definition}

The probabilistic transition function $\mdptrans$ is understood as follows: When the attacker takes an action at the current state, the outcomes of its action will be probabilistic due to the randomized switching of system configurations predefined by the defender's proactive defense strategy and the probabilistic outcome of successfully exploiting the vulnerability. For example, if the system shuffles the IP address, an attack action using the IP address in configuration $i$ will be invalid given the updated system configuration $j$.

We introduce false negative rates for intrusion detectors as follows.
\begin{assumption}
    Given a state-action pair $(s,a)$, if the attack action $a$ is monitored at state $s$, then with probability $1-\epsilon(s,a)$, the attack action will be detected. The value $\epsilon(s, a) \in (0, 1)$ is false negative rate of the detector.
\end{assumption}

Next, we capture the effects of sensors on the attacker's \ac{mdp}.

\begin{definition}[Attacker's MDP given Incomplete Information about Sensor Allocation]
    Given a sensor allocation $(\vec{x},\vec{y})$, the attacker's planning problem is captured by the following \ac{mdp}:
    \[
        \mdp^{\vec{x}} =  \langle Z, A, \mdptrans^{\vec{x}},  \iota,  \calF \rangle ,
    \]
    where $Z, A, \iota, \calF $ are the same as those in the \ac{mdp} without sensors $\mdp$. The transition function $\mdptrans^{\vec{x}}$ is obtained as follows. Consider $(s, i) \in Z$, for each action $a \in A$,
    \begin{enumerate}[(a)]
        \item If $T^j(s' \mid s,a)>0$ and $\vec{x}_{s,j,a}=0$, then we have $ \mdptrans^{\vec{x}}((s', j) \mid (s, i), a) = \mdptrans ((s', j) \mid (s, i), a)  $;
        \item If $T^j(s' \mid s,a) > 0$ and $\vec{x}_{s,j,a}=1$, $\mdptrans^{\vec{x}}((s', j) \mid (s,i), a) = P(i,j) T^j(s' \mid s, a) \epsilon(s, a)$, where $\epsilon(s, a)$ is the state-action dependent false negative rate; In words, if the updated configuration has a detector to monitor the exploitation $(s,a)$ but has a false negative rate $\epsilon(s,a)$,  then the attacker may reach the next state $(s',j)$ at the chance of a detection failure.
        \item If $T^j(s' \mid s,a)$ is not defined, then $\mdptrans^{\vec{x}}((s,j)\mid (s, i), a) =P(i,j)$, which means the defense state changes but no change in the state from the attack graph. 
        \item $\mdptrans^{\vec{x}}(\sink \mid (s,i), a) =\sum_{j\in \topologies} P(i,j) \cdot(1 - \epsilon(s, a))\cdot \vec{x}_{s,j,a}$; In words, the probability of reaching the state $\sink$ is  the probability of getting detected in a configuration at which the intrusion detector is allocated to monitor state-action pair $(s,a)$. 
    \end{enumerate}
\end{definition}

The defender's model of the attack planning problem, described below, is however different due to the use of stealthy sensors.  The following assumption is made.
\begin{assumption}
    A stealthy  sensor has a false negative rate of zero.
\end{assumption}
This assumption is due to the nature of honey patching. It can be relaxed, however, to have false negative rates similar to the treatment for intrusion detector.

\begin{definition}[Defender's MDP given Complete Information about Sensor Allocation]
    Given a sensor allocation $(\vec{x},\vec{y})$, the defender's model of the attack planning problem is captured by the following \ac{mdp}:
    \[
        \mdp^{\vec{x},\vec{y} } =  \langle Z, A,  \mdptrans^{\vec{x},\vec{y}},  \iota,  \calF \rangle ,
    \]
    where $Z, A, \iota, \calF $ are the same as those in the \ac{mdp} without sensors $\mdp$. Consider $(s, i), (s', j) \in Z$, for each action $a \in A$, the transition function $\mdptrans^{\vec{x},\vec{y}}$ is obtained from the transition function $\mdptrans^{\vec{x}}$ in the attacker's \ac{mdp} by letting $ \mdptrans^{\vec{x},\vec{y}}((s', j) \mid (s, i), a) = \mdptrans^{\vec{x}}((s', j) \mid (s, i), a)(1- \vec{y}_{s,j,a})$; and  $\mdptrans^{\vec{x},\vec{y}}(\sink \mid (s, i), a) = \sum_{j\in \topologies} P(i,j) \vec{y}_{s,j,a}$.
\end{definition}

Next, we formulate the defender's value function and the attacker's value function respectively.

By the construction of the attacker's \ac{mdp}, the objective is equivalent to maximizing the probability of reaching the set $\calF$, which is a stochastic shortest path problem~\cite{puterman2014markov}. The optimal attacker's strategy $\pi^\ast$ can be computed by solving the stochastic shortest path problem with the following reward function:
\[
    R(z )=\begin{cases}
        1 & \text{ if $z\in \calF$, } \\
        0 & \text{ otherwise.}
    \end{cases}
\]
This reward function means that a  reward of 1 is received only if the agent reaches a state in $\calF$. In this stochastic shortest path problem, the MDP terminates at an absorbing state. The sink state $\sink$ and $\calF$ are absorbing.

The attacker's perceptual value given the policy $\pi$ and the attacker's \ac{mdp} $\mdp^{\vec{x}}$ is
\begin{multline}
    V_2^\pi (\iota;\vec{x})  = \\ \Expect \left[
        \sum_{k \ge 0}  R (z_k)\mid  z_{k+1} \sim \mdptrans^{\vec{x}}(\cdot \mid z_k,\pi(z_k)), z_0 \sim \iota
        \right]
\end{multline}
where the expectation is taken with respect to the stochastic process induced by policy $\pi$ in $\mdp^{\vec{x}}$ terminating at absorbing states.  That is, $a_k\sim \pi(z_k)$ and $ z_{k+1} \sim \mdptrans^{\vec{x} }( \cdot \mid z_k,a_k)$, for all $k > 0$.

And for the same policy $\pi$, the defender's value is given by
\begin{multline}
    V_1^\pi (\iota;\vec{x}, \vec{y} )  =\\ \Expect \left[
        \sum_{k \ge 0}  R (z_k)\mid  z_{k+1} \sim \mdptrans^{\vec{x}, \vec{y}}(\cdot \mid z_k,\pi(z_k)), z_0 \sim \iota
        \right]
\end{multline}
 where the expectation is taken with respect to the stochastic process induced by policy $\pi$ in $\mdp^{\vec{x},\vec{y}}$ terminating at absorbing states. Note that the terminating time  is perceived differently in the attacker's \ac{mdp} $\mdp^{\vec{x}}$ and the defender's \ac{mdp} $\mdp^{\vec{x},\vec{y}}$ because the attacker cannot observe the stealthy sensors.


The synthesis of sensor allocation is now formulated as a Stackelberg game, in which the defender designs the allocation, in anticipation of the attacker's best response, in the attacker's \ac{mdp} with incomplete information.
\begin{problem}
Let $X\times Y$ be the domains of sensor allocation variables $(\vec{x},\vec{y})$ under the allocation constraints (Assumption~\ref{assume:sensor-constraints}). The  sensor allocation design is a bi-level optimization problem:
\begin{alignat*}{2}
     & \optmins_{(\vec{x}, \vec{y})\in  X\times Y} &       & V_{1}^{\pi^{\ast}}(\iota; \vec{x}, \vec{y})                 \\
     & \optsts                                     & \quad & \pi^{\ast}\in \argmax_\pi   V_{2}^{\pi}(\iota;\vec{x}    ).
\end{alignat*}
\label{pro:subproblem2}
\end{problem}
The bi-level optimization problem is known to be strongly NP-hard~\cite{hansen1992new}. However, we show that due to the special properties of the sensor allocation problem, an optimal solution can be found by reducing it to two single-level \ac{milp} problems. The first one considers optimally allocating intrusion detectors in the absence of stealthy sensors. The second one allocates stealthy sensors given the knowledge of the attacker's best response.

Here, we review \ac{lp} formulation~\cite{de2003linear} for solving the optimal attack policy. Later, we will show how this \ac{lp} formulation facilitates the solution of sensor allocation problems.

Let the optimal value vector be defined by $\vec{v}^\ast=[v^\ast_z]_{z\in Z}$, where $v^\ast_z$ is the probability of reaching $\calF$ from $z$ under the optimal attack policy. We introduce a decision vector $\vec{v}= [v_z]_{z \in Z}$, where $v_z$ is an upper bound on $v_z^\ast$ for each $z\in Z$. Consider the following \ac{lp}:
\begin{alignat}{2}
     & \optmins_{\vec{v}} &       & \sum_{z\in V} c_z v_z  \label{eq:obj_1}                                                            \\
     & \optsts            & \quad & v_z \ge   \sum_{z'\in Z} \calP(z'\mid z,a){v}_{z'}  , \nonumber                                    \\
     &                    &       & \quad \forall a \in A,\ \forall z \in Z, \label{eq:upperbound_1} \\
     &                    &       & v_z = 0,\quad \forall z\in  \{\sink\},  \label{eq:sink}                                            \\
     &                    &       & v_z = R(z),\quad \forall z\in	\calF,\label{eq:final}                                                \\
     &                    &       & v_z \ge 0, \quad \forall z \in Z \label{eq:positive_def_1}
\end{alignat}
where $\vec{c} = [c_z]_{z\in Z}$ is a positive vector, termed as state-relevance weights. The state-relevance weights can be selected to be the initial distribution over the states $Z$. It is shown in~\cite{de2003linear} that any vector $\vec{v}$ that satisfies \eqref{eq:upperbound_1} is an upper bound on the optimal value vector $\vec{v}^\ast$. The objective function is equivalent to minimizing a weighted norm between the upper bound $\vec{v}$ and $\vec{v}^\ast$, given the weight vector $\vec{c} = [c_z]_{z\in Z}$. The solution $\vec{v}$ is shown to be equal to the optimal value vector $\vec{v}^\ast$~\cite{de2003linear}.

From a value function $\vec{v}$, a stochastic attack policy, $\pi: Z\rightarrow \dist{A}$, can be computed as the following equation:
\begin{align}
    \label{eq:policy}
    \pi(a \mid z) = \exp ( ( \qvalue(z, a) - v_z) / \mu),
\end{align}
where $\mu>0$ is a customized temperature. As the $\mu$ goes to $0$, equation~\eqref{eq:policy} recovers hardmax operation. The state-action value function $\qvalue(z,a)$ is  defined by
\begin{equation}
    \label{eq:qvalue}
    \qvalue(z, a) = \sum_{z'\in Z} \mdptrans(z'\mid z,a)v_{z'}.
\end{equation}

\section{Synthesizing the (sub)-optimal sensor  allocation}

\subsection{Step 1: Optimal intrusion detector allocation without stealthy sensors}\label{subsec:ids-allocation}

We first consider the case that the defender only allocates detectors but not stealthy sensors. 
We propose a mixed integer program to solve the optimal intrusion detector allocation strategy as follows. For clarity, we use $x_{s,i,a}$ and $y_{s,i,a}$ to represent $\vec{x}_{s,i,a}$ and $\vec{y}_{s,i,a}$. 
\begin{alignat}{2}
     & \optmins_{\vec{x} \in \mathcal{X}, \vec{v}} &       & \sum_{z\in Z} c_z v_z \label{eq:obj}                                                                     \\
     & \optsts                                     & \quad & v_z \ge  \sum_{(s',j)\in Z}\big( \calP(
     (s', j) \mid z,a){v}_{s',j}  (1- x_{s,j,a})  \nonumber                                                     \\
     &                                             &       & \quad \quad + \calP((s',j)\mid z,a){v}_{s',j}  \cdot x_{s, j ,a} \cdot \epsilon(s,a)\big),\label{eq:upperbound}                   \\
     &                                             &       & \forall a\in A, \forall z= (s, i) \in Z  \setminus (\calF \cup \{\sink\}),  \nonumber                    \\
     &                                             &       & \sum_{(s,a) \in S \times A} x_{s,i,a}\le k, \quad \forall i \in \topologies,	\label{eq:sensor_constraint} \\
     &                                             &       & \eqref{eq:sink}, \eqref{eq:final}, \text{ and } \eqref{eq:positive_def_1},  \nonumber
\end{alignat}
where the domain of variable $\vec{x}$ is $\cal X$ that restricts the allocation to satisfy the constraints in Assumption~\ref{assume:sensor-constraints}.
 When $x_{s,j,a}=1$, the right-hand side of constraint~\eqref{eq:upperbound} is the value given two cases of the next state: The first case is when the attack action is taken but not detected by the intrusion detector. 
In this case of   detection failure, the attack reaches the next state $z'=(s',j)$ from the current state $z = (s, i)$ by taking action $a$ with a probability obtained by the original probability $\mdptrans(z'\mid z,a)$ multiplied with the false negative rate $\epsilon(s,a)$. The second case is when the attack action is taken and detected, the attacker will reach the sink state and the attack terminates. If $x_{s,j,a}=0$, then no intrusion detector is allocated in configuration $j$ to monitor the state-action pair $(s,a)$, then the value is given by $\mdptrans((s',j)\mid z, a) v_{s',j}$.

The constraint~\eqref{eq:upperbound} in the optimization problem is nonlinear due to the product between the variable $v_z$ and the integer variable $x_{s,j,a}$. However, we can introduce new variables to rewrite the problem as an \ac{milp}. Note that the constraint~\eqref{eq:upperbound} is equivalent to
\begin{equation}
    v_{z}\ge\sum_{z'\in Z}\calP(z'\mid z,a)w_{z,a,z'},\quad\forall z\in Z,\ \forall a\in A,\label{eq:new-upperbound}
\end{equation}
where for $z' = (s',j)$,
\begin{equation}
    w_{z,a,z'}=\begin{cases}
        v_{z'} \cdot\epsilon(s,a) & \text{if }x_{s,j,a}=1, \\
        v_{z'}                    & \text{if }x_{s,j,a}=0.
    \end{cases}\label{eq:w_defn}
\end{equation}

Using the big-M method, we can rewrite~\eqref{eq:w_defn} as the following linear constraints:
\begin{subequations}
    \label{eq:w_v}
    \begin{align}
        w_{z,a,z'}-  v_{z'}\cdot\epsilon(s,a)  & \le M\cdot(1-x_{s,j,a}),\label{eq:w_v_1} \\
        w_{z,a,z'}-  v_{z'} \cdot\epsilon(s,a) & \ge m\cdot(1-x_{s,j,a}),\label{eq:w_v_2} \\
        w_{z,a,z'} -  v_{z'}                   & \le M\cdot x_{s,j,a},\label{eq:w_v_3}    \\
        w_{z,a,z'} -  v_{z'}                   & \ge m\cdot x_{s,j,a},\label{eq:w_v_4}
    \end{align}
\end{subequations}where $M$ and $m$ are constants to be defined shortly. When $x_{s,j,a}=1$, the constraints~\eqref{eq:w_v_1} and~\eqref{eq:w_v_2} together recover $w_{z,a,z'}= v_{z'}\cdot\epsilon(s,a)$, whereas the constraints~\eqref{eq:w_v_3} and~\eqref{eq:w_v_4} become non-binding as long as $M$ and $m$ are chosen appropriately. For this problem, it is not difficult to verify that it suffices to choose $M=1$ and $m=-1$. A similar argument can be made for the case when $x_{s,j,a}=0$. The final form of the \ac{milp} is given as follows:
\begin{alignat*}{2}
     & \optmins_{\vec{x} \in \mathcal{X}, \vec{v}} &       & \sum_{z\in Z} c_z v_z \nonumber                                                                                                        \\
     & \optsts                                     & \quad & \eqref{eq:sink}, \eqref{eq:final}, \eqref{eq:positive_def_1}, \eqref{eq:sensor_constraint}, \eqref{eq:new-upperbound}, \eqref{eq:w_v}, \\
     &                                             &       & w_{z,a,z'}\ge 0, \quad \forall z \in Z,\ \forall a \in A,\ \forall z' \in Z.
\end{alignat*}

\subsection{Step 2: Optimal stealthy sensor allocation for a fixed detector allocation}

Next, we allocate a bounded number of stealthy sensors given the attacker's policy $\pi^\ast$, calculated from the attacker's \ac{mdp} $\mdp^{\vec{x}}$.
In addition, we introduce decision variables $\vec{v}=[v_z]_{z\in Z}$, where $v_z$ is the optimal attack success rate given both intrusion detector and stealthy sensors and new decision variables $\vec{q} = [q_{z,a, z'}]_{(z, a, z') \in Z \times A \times Z}$. 
We propose another \ac{milp} for computing the optimal stealthy sensor allocation strategy:
\begin{alignat}{2}
     & \optmins_{\vec{q}, \vec{v}, \vec{y}\in \mathcal{Y}} &       & \sum_{z\in Z} c_z v_z  \label{eq:obj_2}                                                                     \\
     & \optsts                                             & \quad &
    v_z = \sum_{(a, z') \in A \times Z} q_{z, a, z'},\quad \forall z \in Z, \label{eq:policy-eval}                                                                               \\
     &                                                     &       & q_{z,a,z'} \le M \cdot( 1 - y_{s,i,a}), \label{eq:honey-constraint-1}                                       \\
     &                                                     &       & \mdptrans(z' \mid z,a)\pi^\ast(z,a)v_{z'} - q_{z,a,z'} \ge m \cdot y_{s,i,a}, \label{eq:honey-constraint-2} \\
     &                                                     &       & \mdptrans(z' \mid z,a)\pi^\ast(z,a)v_{z'} - q_{z,a,z'} \le M \cdot y_{s,i,a}, \label{eq:honey-constraint-3} \\
     &                                                     &       & q_{z, a, z'} \ge 0, \label{eq:positive_def_2}                                                               \\
     &                                                     &       & \forall a \in A, \forall z = (s, i) \in Z, \forall z' = (s',j) \in Z, \nonumber                             \\
     &                                                     &       & \sum_{s, a} y_{s,i,a} \le  h,\quad \forall i \in \topologies, \label{eq:limited-stealthy}                   \\
     &                                                     &       & \text{and} \eqref{eq:sink}, \eqref{eq:final}, \eqref{eq:positive_def_1} \nonumber,
\end{alignat}
where $M=1$ and $m=-1$ are constants. The domain of variable $\vec{y}$ is $\cal Y$ that restricts the allocation to satisfy the constraints in Assumption~\ref{assume:sensor-constraints}. For this optimization problem, we aim to minimize the weighted sum of attack success rate $\vec{v}$ in~\eqref{eq:obj_2}. Note that if the weights $\vec{c} = [c_z]_{z\in Z}$ are chosen to be the initial state distribution, the objective function in \eqref{eq:obj_2} is equivalent to minimizing the attack success rate given the initial distribution.

Constraint \eqref{eq:policy-eval} enforces that the state value $v_z$ is the summation over state-action-state value $q_{z, a, z'}$ for all actions $a \in A$ and next states $z'\in Z$. Constraint \eqref{eq:honey-constraint-1} means that if $y_{s,i, a}=1$, then the state-action-state value $q_{z,a, z'} = 0$ as the attacker will be detected.
If $y_{s,i, a}=0$, constraints \eqref{eq:honey-constraint-2} and \eqref{eq:honey-constraint-3} enforce
\begin{equation}
    \mdptrans(z' \mid z,a)\pi^\ast(z,a)v_{z'} = q_{z, a, z'}.
    \label{eq:honey-constraint-23}
\end{equation}
Substituting $q_{z,a,z'}$ into \eqref{eq:policy-eval}, we have policy evaluation of $\pi^\ast$ given the stealthy sensors and intrusion detectors allocation. In the end, we consider finite number of stealthy sensors constrained by inequality \eqref{eq:limited-stealthy}. Constraint \eqref{eq:positive_def_2} means the state-action-state values are non-negative. 

\section{Case Study}
\label{sec:case-study}
\begin{figure}[htbp]
    \centering
    \includegraphics[width=0.9\linewidth]{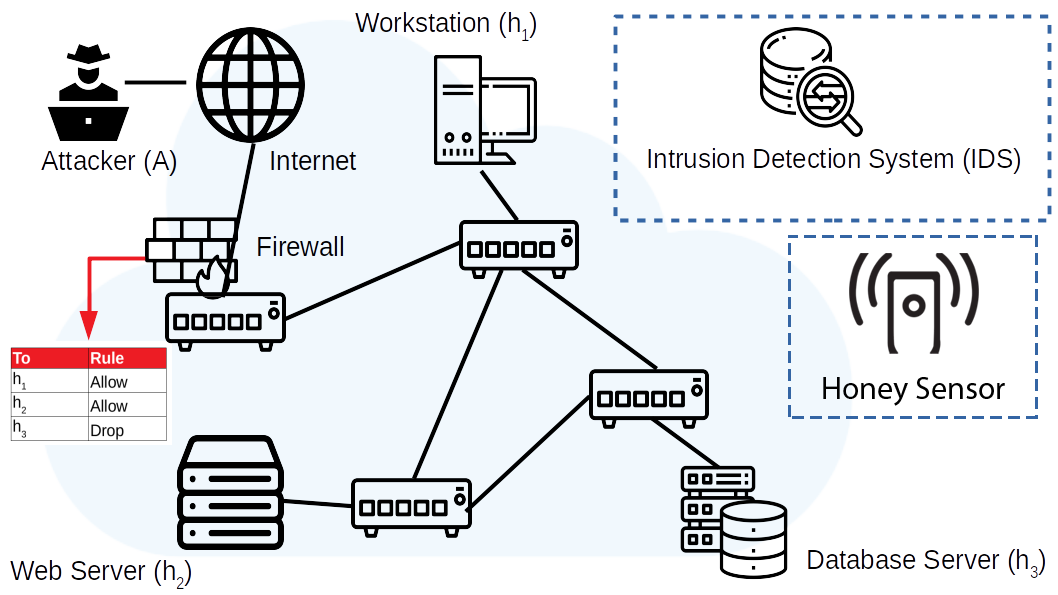}
    \caption{Network example.}
    \label{fig:network-system}
\end{figure}

To illustrate the effectiveness of the proposed method, we consider an example of a cyber system shown in Fig.~\ref{fig:network-system} inspired by~\cite{yoonAttackGraphbasedMoving2020}. The system has three hosts: the workstation $h_1$ handles users' requests, the webserver $h_2$ handles web service requests, and the database server $h_3$ houses critical data such as personal credentials. In addition, there are a few network security functions, such as firewall, intrusion detectors, and stealthy sensors available to be deployed in the network. The firewall divides hosts into hosts that internal entities can access and hosts that outside entities can access. In this example, $h_1$ and $h_2$ can be accessed by outside entities, and $h_3$ can only be accessed by internal entities. The attacker is initially outside the network system, and the goal is to acquire root privilege on host $h_3$.

\begin{figure}[htbp]
    \centering
    \resizebox{0.7\linewidth}{!}{%
        \begin{tikzpicture}[->,>=stealth',shorten >=1pt,auto,node distance=5cm,scale=1, semithick, transform shape]
            \tikzstyle{every state}=[fill=black!10!white]
            \node[initial, state]   (0)                 {$0$};
            \node[state]            (1) [right of=0]    {$1$};
            \path[->]
            (0) edge[loop above]    node    {$0.3$}     (2)
            (0) edge[bend left]     node    {$0.7$}     (1)

            (1) edge[loop above]    node    {$0.6$}     (1)
            (1) edge[bend left]     node    {$0.4$}     (0)
            ;
        \end{tikzpicture}
    }
    \caption{Two-state proactive \ac{mtd} strategy.}
    \label{fig:transducer}
\end{figure}
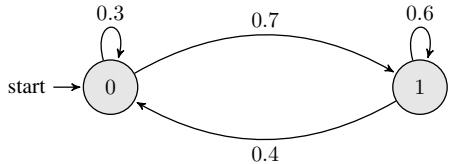

We equip this network with a proactive redundancy-based \ac{mtd} strategy; that is, we have replicas of \ac{os} for network components, and the network configuration is updated dynamically. More specifically, the hosts $1$ and $2$ probabilistically switch between default \ac{os}s and backup \ac{os}s \footnote{The information about the default \ac{os}s and backup \ac{os}s, along with all the vulnerabilities, \ie, attack actions, can be found in \url{https://bit.ly/3xWxdDa}.} This proactive \ac{mtd} strategy is captured by a Markov Chain shown in Fig.~\ref{fig:transducer}.

The Markov Chain can be understood as follows: at the state $0$, the network \ac{mtd} controller either switches to backup \ac{os}s with probability $0.7$ or stay with the default \ac{os}s with probability $0.3$; at the state $1$, the network system switch back to default \ac{os}s with a probability $0.4$ or stay with the backup \ac{os}s with probability $0.6$. In this example, the finite defender states $0$ and $1$ have one-to-one mappings to the set of network configurations (default and backup).

\begin{figure}[htbp]
    \centering
    \resizebox{0.9\linewidth}{!}{%
        \begin{tikzpicture}[->,>=stealth',shorten >=1pt,auto,node distance=4cm,scale=0.8, semithick, transform shape,state/.style={ellipse, draw,fill=black!10!white, minimum size=3em}]
            \tikzstyle{every state}=[fill=black!10!white]
            \node[state]   (0)                          {$(\text{A}, 0)$};
            \node[state]            (1) [right of=0]    {$(\text{A}, 1)$};
            \node[state]            (2) [below of=0]    {$((h_1, \text{user}), 0)$};
            \node[state]            (3) [above of=0]    {$\sink$};
            \node[state]            (4) [below of=1]    {$((h_2, \text{root}), 1)$};

            \path[->]
            (0) edge[bend right]    node[right] {$w_1, 0.21$}     (3)
            (0) edge[loop above]    node        {$w_1, 0.027$}    (0)
            (0) edge                node        {$w_1, 0.7$}      (1)
            (0) edge                node        {$w_1, 0.063$}    (2)
            (0) edge[in=135,out=165,looseness=8]    node      {$ub_1, 0.3$}  (0)
            (0) edge[bend left]                     node      {$ub_1, 0.7$}  (1)
            (0) edge[in=195,out=225,looseness=8]    node    {$ws_3, 0.3$}    (0)
            (0) edge[bend left]                     node    {$ws_3, 0.49$}   (3)
            (0) edge[bend right]                    node    {$ws_3, 0.147$}  (1)
            (0) edge                                node    {$ws_3, 0.063$}  (4)
            ;
        \end{tikzpicture}
    }
    \caption{A fragment of MDP constructed from Def~\ref{def:attackMDP}.}
    \label{fig:part-product-mdp}
\end{figure}
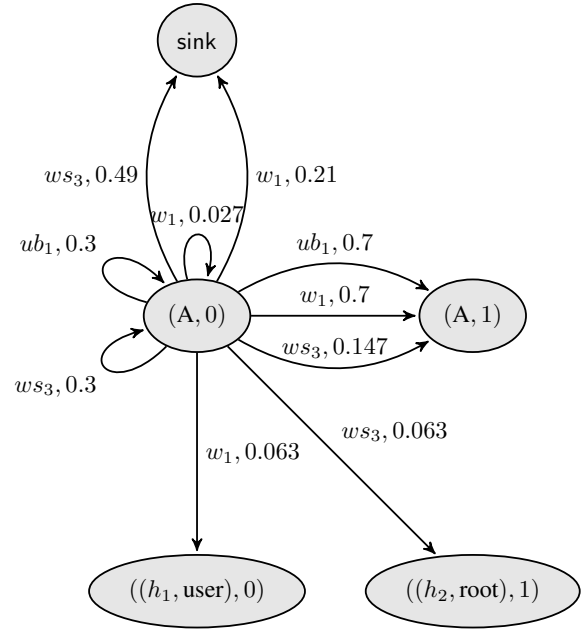

For each network configuration, we generate its corresponding host-based attack graphs~\cite{hewett2008host}   based on the vulnerabilities from \ac{cvss}~\cite{CVSSV3Specification}. Note that state $(h_2, \text{user})$ is not reachable in the attack graph for state $0$ and thus omitted from the figure. Given the attacker's objective is to reach root privilege in host $h_3$, the set of goal states in the attack graphs is $\{(h_3,\text{root})\}$ for both attack graphs. The set of final states in the attacker's planning problem (Def.~\ref{def:attackMDP}) is $\{((h_3,\text{root}), 0), ((h_3,\text{root}), 1)\}$. To illustrate the attack planning problem, we plot a fragment of the attacker's \ac{mdp} in Fig.~\ref{fig:part-product-mdp}. The initial state is $(\text{A}, 0)$, and the attacker can take action $w_1$ to reach state $((h_1, \text{user}),0)$ with probability $0.063$, which is calculated based on the product of three quantities: \begin{inparaenum}[1)]\item the probability of staying in configuration $0$ ($0.3$); \item the probability of exploiting the vulnerability $w_1$ successfully ($0.7$); \item the false negative rate $\epsilon = 0.3$ for the intrusion detector is deployed in $((h_1, \text{user}), 0, w_1)$ but missed the detection\end{inparaenum}. We assume for each state except for the target $(h3, root)$, for each attack action, an intrusion detector or a stealthy sensor can be allocated to monitor that state-action pair.

\begin{figure}[htbp]
        \centering
        \begin{minipage}{0.8\linewidth}
            \begin{tikzpicture}
                \node (img)  {\includegraphics[width=\linewidth, trim={25 20 0 0},clip]{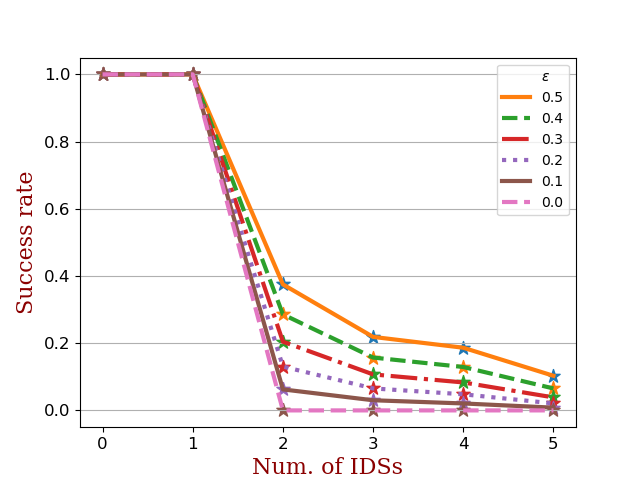}};
                \node[below=of img, node distance=0cm, yshift=1cm,font=\color{red}] {Num. of Intrusion Detectors};
                \node[left=of img, node distance=0cm, rotate=90, anchor=center,yshift=-0.7cm,font=\color{red}] {Success rate};
            \end{tikzpicture}
        \end{minipage}%
        \caption{The number of intrusion detectors versus the attack success rates under different false negative rates $\epsilon$.}
        \label{fig:ids-vs-epsilon}
    \end{figure}

In the first step, we solve the optimal intrusion detector allocation problem, with varying upper bounds on the number of deployable intrusion detectors and varying false negative rates. We assume the same false negative rates for all intrusion detectors to illustrate how the false negative rate affects the effectiveness of defense. Note that the algorithm allows different intrusion detectors with different false negative rates. Fig~\ref{fig:ids-vs-epsilon} summarizes the results. When the false negative rate is fixed, the attack success rates are monotone and non-increasing as the number of intrusion detectors increases. That means with the more intrusion detectors the system can deploy, the attacker has less chance to achieve the target because although he observes intrusion detectors, due to the randomization it cannot always evade intrusion detectors. When the number of intrusion detectors is fixed, the success rates are monotone and non-increasing as the false negative rate decreases. When the false negative rate $\epsilon = 0.3$ and the number of intrusion detectors is $4$, intrusion detectors should be placed at $\{(\text{A}, r_1), (\text{A}, w_1), ((h_1,\text{root}), b_3), ((h_2,\text{root}), b_3)\}$ at state $0$ and $\{(\text{A}, ws_3), ((h_1,\text{user}), b_3), ((h_2,\text{root}), b_1)$,\\$((h_2,\text{root}), b_3)\}$ at state $1$.

    When the number of intrusion detectors is $1$, for all $\epsilon$ ranging from $0$ to $0.5$, we show that the attacker can reach the target state $(h_3, \text{root})$ with probability $1$. The solution suggests placing intrusion detectors at $((h_2, \text{root}), b_3)$ at $0$ and $((h_1, \text{user}), b_3)$ at $1$, but one intrusion detector at each configuration is not sufficient to block alternative attack actions. For example, when the configuration is at $0$ and the attacker reaches the state $(h_2, \text{root})$, the intrusion detector is located at $((h_2, \text{root}), b_3)$, but the attacker take action $b_1$ to reaches the target with probability $1$.

\begin{figure}[htbp]
        \centering
        \begin{minipage}{0.8\linewidth}
            \begin{tikzpicture}
                \node (img)  {\includegraphics[width=\linewidth, trim={30 20 0 0},clip]{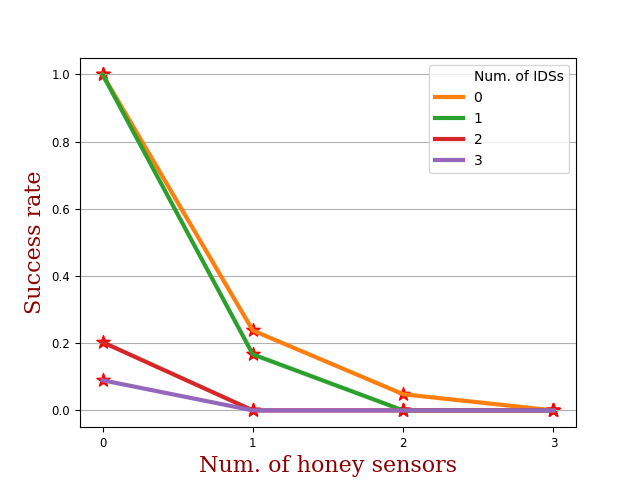}};
                \node[below=of img, node distance=0cm, yshift=1cm,font=\color{red}] {Num. of Stealthy Sensors};
                \node[left=of img, node distance=0cm, rotate=90, anchor=center,yshift=-0.7cm,font=\color{red}] {Success rate};
            \end{tikzpicture}
        \end{minipage}%
        \caption{The number of stealthy sensors versus the attack success rates, where the number of intrusion detectors is $k \in \{0, 1, 2, 3\}$, and false negative rate is $\epsilon=0.3$.}
        \label{fig:ids-vs-honey}
    \end{figure}

    After solving the optimal intrusion detector allocation, we synthesize the optimal stealthy sensor allocation strategy. We first extract the optimal attacker's policy according to \eqref{eq:policy} and \eqref{eq:qvalue}, where the temperature $\mu$ is $0.1$. We vary the number of intrusion detectors and the number of stealthy sensors and fix the false negative rate $\epsilon = 0.3$. Fig.~\ref{fig:ids-vs-honey} summarizes the attack success rates and indicates that, if we fix the number of intrusion detectors and the corresponding policy, the success rates are monotone and non-increasing as the number of stealthy sensors increases.

    Furthermore, we compare two cases with a false negative rate $\epsilon = 0.3$: (a) one intrusion detector and one stealthy sensor; (b) two intrusion detectors. For case (a), the attack success rate is $0.167$; for case (b), the attack success rate is $0.205$. This comparison shows that, for the same number of sensors, deploying stealthy sensors is more effective (with $18.5\%$ reduction in the attack success rate) because first, the stealthy sensor has zero false negative rate, and second, the attacker cannot observe these stealthy sensors and plan to evade them. We consider a case when false negative rate $0.3$, and there are $2$ intrusion detectors and $1$ stealthy sensor available. The solution suggests we deploy intrusion detectors at $\{(\text{A}, w_1), (\text{A}, r_1)\}$ and stealthy sensor at $\{(\text{A}, r_1)\}$ at $0$; we deploy intrusion detectors at $\{(\text{A}, w_1), (\text{A}, ws_3)\}$ and stealthy sensor at $\{(\text{A}, r_1)\}$ at $1$ \footnote{We provide the constructed attack graphs and solutions for all intrusion detectors and stealthy sensor allocations in the following link: \url{https://bit.ly/3zwHrtm}.}.

    The \acp{milp} are solved using the Python-MIP package with Gurobi 9.1.2 on a Windows 10 machine with Intel(R) Xeon (R) E5-1607 v3 CPU and 16~GB RAM. The average computational time of intrusion detectors allocation is $0.72$~s, and the average computational time of stealthy sensors allocation is $0.58$~s.

\section{Conclusions}\label{sec:conclusion}
For an attacker compromising a cyber system equipped with a proactive \ac{mtd} mechanism, we developed a formal method-based modeling framework to capture the attacker's planning problems and synthesis algorithms for optimally allocating sensors that minimize the attack success rate. We specifically considered two types of sensors: intrusion detectors that are observable to the attacker and stealthy sensors that are not observable to the attacker. The experiment results demonstrate the combined benefit of \ac{mtd}, intrusion detection, and deception. In our future work, the following extensions will be investigated: First, our current formulation to allocate stealthy sensors assumes that the attacker is unaware of the use of cyber deception. It remains open to investigate the design of stealthy sensor allocation given deception-aware attacker. Second, the current formulation considers one-time interaction. To mitigate persistent attackers, one must consider that the attacker may learn the deployment of stealthy sensors from past interactions and improve its attack policy.  Lastly, we assume a powerful attacker who can observe the defender's states. In practice, if the defender's states are different from network configurations, then the attacker may not be able to construct the defender's \ac{mtd} strategy or observe partially the states in the attack planning problem. It is of practical interest to investigate the sensor allocation against attackers with partially observations.


\bibliographystyle{ieeetr}
\bibliography{refs}

\end{document}